\documentclass{article}
\usepackage{icmc2024template}
\usepackage{times}
\usepackage{ifpdf}
\usepackage{soul}
\usepackage[english]{babel}

\def\papertitle{Collaboration Between Robots, Interfaces and Humans: \\ Practice-Based and Audience Perspectives}
\def\firstauthor{Anna  Savery}
\def\secondauthor{Richard Savery}
\def\thirdauthor{Third Author}

\newif\ifpdf
\ifx\pdfoutput\relax
\else
   \ifcase\pdfoutput
      \pdffalse
   \else
      \pdftrue
  \fi
\fi

\ifpdf 
  \usepackage[pdftex,
    pdftitle={\papertitle},
    pdfauthor={\firstauthor, \secondauthor, \thirdauthor},
    bookmarksnumbered, 
    pdfstartview=XYZ 
   ]{hyperref}

  \usepackage[pdftex]{graphicx}
  \graphicspath{{./figures/}}
  \DeclareGraphicsExtensions{.pdf,.jpeg,.png}

  \usepackage[figure,table]{hypcap}

\else 
  \usepackage[dvips,
    bookmarksnumbered, 
    pdfstartview=XYZ 
  ]{hyperref}  

  \usepackage[dvips]{epsfig,graphicx}
  \graphicspath{{./figures/}}
  \DeclareGraphicsExtensions{.eps}

  \usepackage[figure,table]{hypcap}
\fi

\hypersetup{
    colorlinks,%
    citecolor=black,%
    filecolor=black,%
    linkcolor=black,%
    urlcolor=black
}

\title{\papertitle}

\twoauthors
  {1.5in}
  {\firstauthor} {University of Technology Sydney \\  %
    {\tt \href{mailto:anna.Savery@uts.edu.au}{anna.Savery@uts.edu.au}}}
  {\secondauthor} {Macquarie University \\  %
    {\tt \href{mailto:richard.savery@mq.edu.au}{richard.savery@mq.edu.au}}}

\begin{document}
\capstartfalse
\maketitle
\capstarttrue
\begin{abstract}
This paper provides an analysis of a mixed-media experimental musical work that explores the integration of human musical interaction with a newly developed interface for the violin, manipulated by an improvising violinist, interactive visuals, a robotic drummer and an improvised synthesised orchestra. We first present a detailed technical overview of the systems involved including the design and functionality of each component. We then conduct a practice-based review examining the creative processes and artistic decisions underpinning the work, focusing on the challenges and breakthroughs encountered during its development. Through this introspective analysis, we uncover insights into the collaborative dynamics between the human performer and technological agents, revealing the complexities of blending traditional musical expressiveness with artificial intelligence and robotics. To gauge public reception and interpretive perspectives, we conducted an online survey, sharing a video of the performance with a diverse audience. The feedback collected from this survey offers valuable viewpoints on the accessibility, emotional impact, and perceived artistic value of the work. Respondents' reactions underscore the transformative potential of integrating advanced technologies in musical performance, while also highlighting areas for further exploration and refinement.

\end{abstract}
%

\begin{figure*}[h]
	\centering
		\includegraphics[width=14cm]{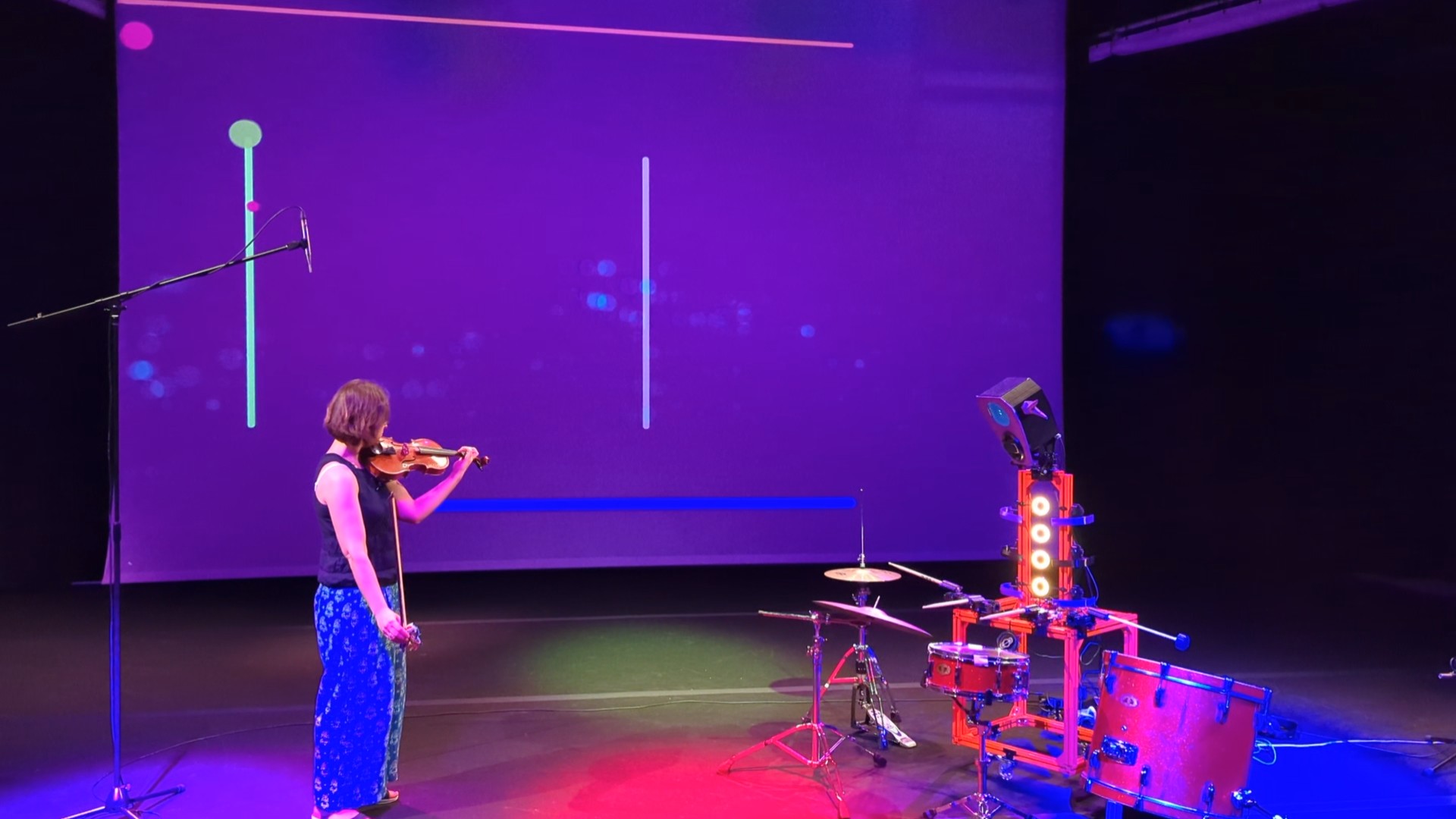}
	\caption{Performance of the combined work}
	\label{fig:combined}
\end{figure*}

\section{Introduction}\label{sec:introduction}
In music technology research, collaboration between independently developed new interfaces and robotic musicianship remains a relatively under-explored domain \cite{sullivan2018stability}. Despite the transformative potential, collaborations within these areas are infrequent, possibly due to the challenges inherent in merging disparate technological systems and the often limited resources allocated for such interdisciplinary endeavors. In work on new musical interfaces 78\% of practitioners fulfil the role of designer, composer and performer \cite{morreale2018nime}.

However a key component of musical creation and appreciation lies within communal activity—ensemble performances, composer-performer collaborations, and the shared experiences between performers and their audiences \cite{small1998musicking}. This inherent collectivity in music poses unique challenges within the context of new music technologies, where the integration of diverse technologies is not always straightforward, and the development of music for new technological mediums can be particularly complex. 

The introduction of novel interfaces and robotic musicians into traditional performance settings raises questions regarding their impact on both performers and audiences. How do performers adapt to and integrate their technologies into the practice of others, and how does the audience perceive and engage with these technologically augmented performances? This paper aims to develop new understanding into these questions by exploring a case study on integration of robotic musicianship and innovative interface design within a mixed-media experimental music framework. 

This paper describes and analyzes a specific experimental musical performance that integrates a novel violin interface, human and AI improvisation, interactive visuals, and robotics (see Figure \ref{fig:combined}). This grouping offers a unique area to explore the tension between traditional and technological modalities in music. Our documentation and discussions aim to dissect these dynamics, shedding light on the challenges and inherent potential in integrating cutting-edge technology with human artistry in musical performance. Our analysis is grounded in the case study provided by the referenced performance, serving as a detailed example of the broader implications of technology in the arts.

We start with a technical description of each system, detailing the design and implementation of \textit{Keirzo}, the robotic musician, followed by a breakdown of the violin interface and interactive visuals developed for this piece. In our description, we examine both, the hardware and software parameters of each technology, highlighting the intention behind their development and explaining their functionalities. We then present our views on what we consider to be significant discussion points associated with this practice-led research project. These points include our perspectives on embodiment and disembodiment of sound, human musician interaction with a robotic musician collaborator, staging and physicality of the referenced work and strategies for designing the visualisation.

The final section of this paper presents an analysis of our piece from an audience perspective, with data derived from an online study conducted using Prolific. Participants were asked to view a video of the performance with no prompt and then described their reactions. Audience perspectives from our survey predominantly echoed our own discussion points as well as providing fresh insight and view points which will likely influence and shape our future compositions. One of the key takeaways is the importance in audience accessibility and education when presenting new experimental works, specifically mixed-media pieces involving robotics and AI, where audience members may experience feelings of prejudice and hostility unless full transparency is displayed. After analysing the survey data, we have come to the conclusion that incorporating these strategies into future compositions will result in better ways of communicating artistic intentions and create a more inclusive space for audiences unfamiliar with new technologies.

A demonstration video can be seen online.\footnote{\href{https://drive.google.com/drive/folders/15DQJGOxFW5TuLm3_TYA5qQIcdZxPTK4-?usp=sharing}{https://bit.ly/violinrobot}}

\section{Related Work}

\subsection{Instrument Augmentation}


Bowed string instruments have been extensively enhanced with a wide range of sensor-based technologies~\cite{overholt2014advancements}. The \textit{Hyperinstruments} research project, founded by Tod Machover 1981~\cite{monteverdisound} has produced a series of augmented instruments aimed at extending an existing skill set of an expert practitioner through sophisticated gesture analysis and bespoke software environments~\cite{bevilacqua:hal-01161349, machover1994interactive}. Notable projects include the \textit{Hypercello}, developed for Yo-Yo Ma's performance of Machover's \textit{Begin Again Again...}~\cite{paradiso1997musical}, the \textit{Hyperviolin}, originally developed for Ani Kavafian~\cite{overholt2014advancements} followed by a more recent version, updated in association with Joshua Bell~\cite{young_hyperbow_2002}. Both versions used iterations of Young's \textit{Hyperbow}~\cite{rasamimanana2004gesture}, with embedded sensors allowing real-time gesture and bow stroke analysis.

Kimura's \textit{Augmented Violin Glove}~\cite{kimura_extracting_nodate} builds on the \textit{Hyperinstruments} research by incorporating gesture analysis technology into a custom designed glove. Built in collaboration with IRCAM researchers, Kimura's \textit{Augmented Violin Glove} allows an expert violinist to extend their creativity using a non-invasive approach~\cite{kimura_extracting_nodate}. Veering away from the \textit{Hyperinstruments} project and natural gesture capture, Overholt's \textit{Overtone Violin}~\cite{overholt2005overtone} and Ko and Oehlberg's \textit{TRAVIS II}~\cite{ko2020construction} offer the violinist an opportunity to learn new extended techniques in order to master these augmented instruments. Both violins retain the essence of traditional acoustic instruments with the addition of real-time interaction tools in the form of attached sensors and other instrument modifications.

Other new musical interfaces that aim to extend an existing skill-set of an expert practitioner with sensor-based technologies include Impett's \textit{Meta-Trumpet}~\cite{impett1994meta} and more recently, Reid's \textit{MIGSI}~\cite{reid2016minimally}. Although differential in methods and design, these sensor-enhanced musical interfaces offer expert practitioners opportunities to extend their performance practice through unique and creative approaches to technological applications.

\subsection{Robotic Musicianship}
Robotic musicianship is an  interdisciplinary field combining music, computer science, performance studies, AI, and mechatronics (the physical creation of robotics) \cite{frid2023musical}. The field of robotic musicianship aims to develop robots with musical abilities, such as to listen and improvise with human performers \cite{savery2022robotics}. Music styles explored in robotic musicianship, including jazz improvisation and free style hip hop interaction, offer communicative dialogue-based mediums that require extensive domain knowledge \cite{savery2020shimon}. Robotic musicianship is often considered a separate field to musical mechatronics, which instead focuses on actuation or control of instruments, without necessarily developing agency or a robot persona \cite{yang2020mechatronics}. Robotic musicians can be compared in different ways, for example, Kemper classifies robots into anthropomorphic representation of robots, their ability to produce nuanced sounds, the level of control over their musical expression, and their overall musical output \cite{kemper2021locating}.

\section{Violin Interface}\label{sec:introduction}
\begin{figure}[h]
	\centering
		\includegraphics[width=\columnwidth]{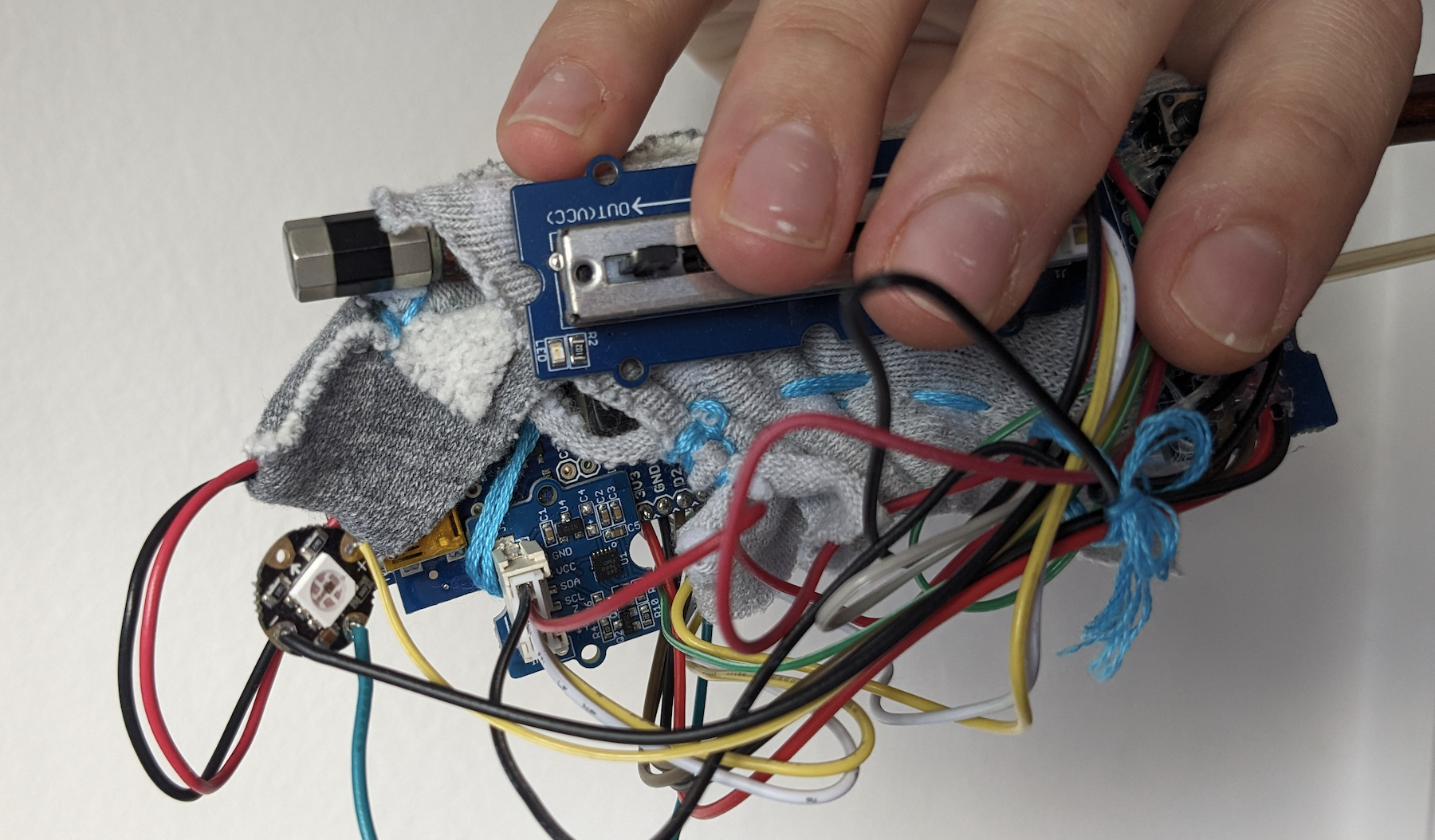}
	\caption{Violin bow interface used in this piece}
	\label{fig:current.png}
\end{figure}

In this piece, the violinist's bow is augmented with a custom-designed musical interface that is used to interact with both, Keirzo and the visuals. 
This interface has been in development since January 2021 \cite{savery2023enhancing} and is designed to be minimally invasive, with all the sensors attached to a stretchy cotton fabric that slides on and off the violin bow (Figure~\ref{fig:current.png}). This design approach prevents any permanent modifications to the existing bow.
Data from the interface is sent wirelessly using the ZigBee protocol~\footnote{\href{https://www.digi.com/products/embedded-systems/digi-xbee}{ZigBee Wireless Mesh Networking}}. This is achieved by using an XBee radio module within the interface and an XBee receiver unit attached to a computer via USB.
The interface is fitted with an accelerometer sensor that tracks the bow movements in real-time, an RGB NeoPixel, three push-buttons and a potentiometer slider. These sensors are wired into an Arduino Fio which is powered by a small lithium polymer 3V battery, allowing for freedom of right-hand movement without cables or heavy battery packs.

The virtual component of this interface uses Max/MSP and Processing software environments, to create an audio and visual output unique to each composition project. By utilising this interface, an expert violinist is able to interact with a custom-made system that integrates both, natural gesture tracking and extended techniques through finger manipulation of the various sensors.

\begin{figure}[]
	\centering
		\includegraphics[width=\columnwidth]{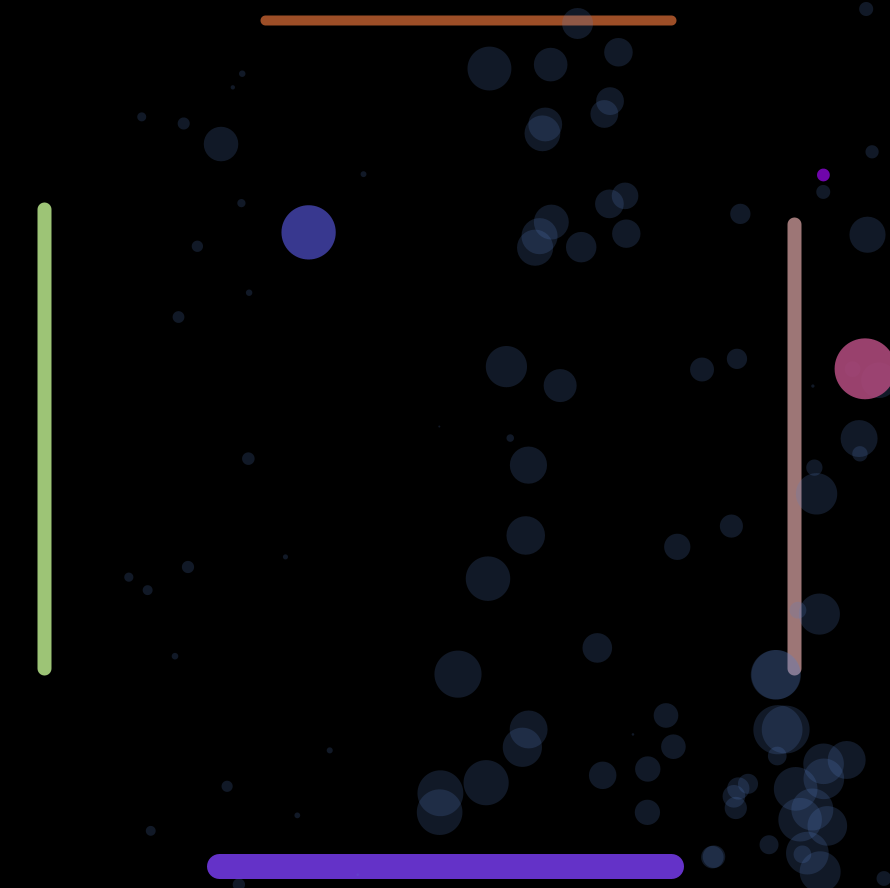}
	\caption{Interactive visuals capturing violinist's bow gestures and Keirzo's drum hits}
	\label{fig:visuals2.png}
\end{figure}

\section{Visuals}\label{sec:introduction}
The visuals for this work were developed using the Processing software environment, exploring three concepts:
\begin{itemize}
    \item Real-time bow gesture tracking using particle systems
    \item Real-time visualisation of \textit{Keirzo's} drum hits
    \item Simple collision behaviour between three ellipses and the lines representing \textit{Keirzo's} drum hits 
\end{itemize}

The particle system follows the violin bow using an algorithm that maps bow placement in relation to horizontal motion and positioning within each string using the accelerometer sensor x and y coordinates. This mapping approach is effective in visualising the horizontal and vertical motion of the violinist's bow in real-time. Further mappings utilised acceleration values to create a relationship between fast and slow bow strokes, incorporating a wind force to mimic the speeding up and slowing down of the system in relation to the music.

Max/MSP is used to receive both, the interface data and midi values from Keirzo. Each time Keirzo's drum sticks come into contact with one of his four drums, a value is sent into the violinist's Max patch. These are numbered 1-4 and correspond to the four lines shown within the Processing canvas window. For every hit, the lines move forward and sideways using randomised but constrained values so that each line will not leave the dimensions of the window. Once a line travels to the end of the canvas window dimensions, it reverses its direction. 

The lines are also interacting with the three larger ellipses using simple collision detection. The spheres bounce off the lines as well as each other. These collisions affect the changes in colour and size of each ellipse, creating a playful augmentation to the physical performance presence (Figure~\ref{fig:visuals2.png}.

\begin{figure}[h]
	\centering
		\includegraphics[width=\columnwidth]{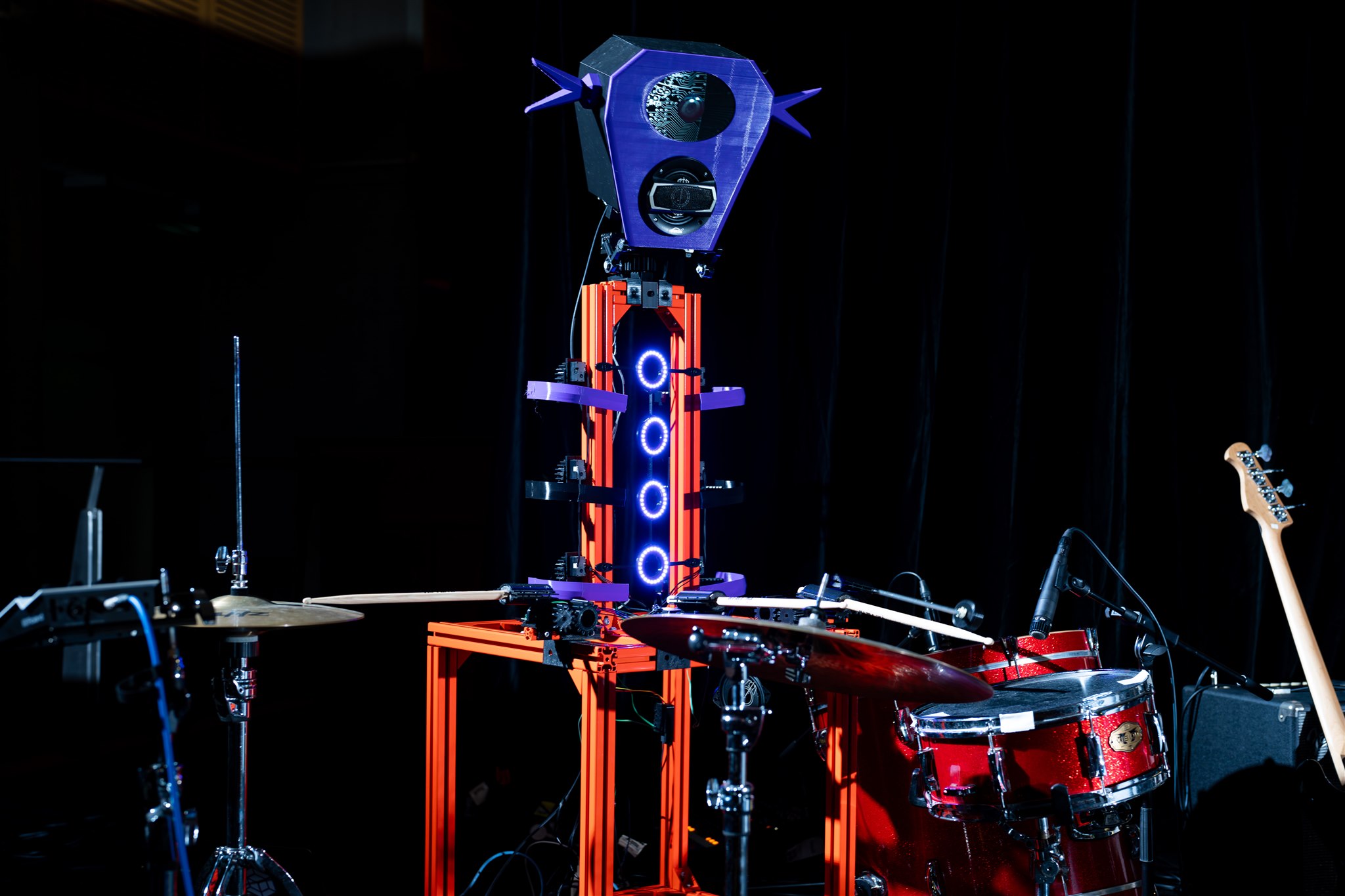}
	\caption{Keirzo the Robotic Musician in Performance}
	\label{fig:perf.jpeg}
\end{figure}

\section{Keirzo}\label{sec:introduction}

In this work, we utilize Keirzo, a custom-built robot with capabilities in drumming and improvisation \ref{fig:perf.jpeg}. Keirzo is specifically designed to engage with musicians and other humans. The system has been constructed to facilitate extensive live performances and emphasises human interaction. Keirzo is contains five interactive components, drumming arms, a rib cage, LED rings, head movement and an eye display. 

The drumming uses four dynamixel XH430-W210-R servomotors, each manipulating a drumstick through. These arms are intentionally designed with an emphasis on visual transparency of their function, opting for exaggerated, large-scale movements. This design choice prioritizes the visibility of the robot's actions over mechanical efficiency, deliberately placing the arms in a position  on the robot and away from drum set rather than integrated directly onto the drum set. 

Encircling the central structure of Keirzo is what is referred to as the 'rib cage.' This element does not contribute directly to the musical output but is essential for the robot's visual expressiveness. It is  designed to simulate the appearance of breathing and other human-like gestures, aiming to enhance the robot's ability to communicate movement and engage with onlookers and fellow performers. The LED rings are used to demonstrate pulse, and placed in custom enclosures for 4 Neopixel rings.

The robot's head is equipped with three degrees of freedom, enabling it to convey beats and musical patterns through motion. This movement serves as a non-verbal cueing system, allowing Keirzo to signal changes in the musical arrangement or to synchronize with other musicians. Finally, the 'eye display' of Keirzo responds in real-time to audio inputs. This display is powered by a custom-programmed interface developed in Max/MSP/Jitter, offering a highly flexible platform for visual feedback that can be adapted to various performance contexts.

For this robot performance Keirzo's drumming was integrated with a deep learning based, virtual synthesized orchestra. The rhythm of the orchestra was generated by a rule-based algorithms controlled by the violin. There were three modes of performance, in-tempo staccato backings, longer sustains, and short tremolo interjections. Each mode placed beats on a grid, with predefined probabilities for each grid space. The violinist was able to choose which mode with their interface, as well as the density of texture and the tempo of the grid (between 80 and 200 bpm). In addition to the backing their were generated melodies that were created using the set density and the z angle of the bow, with movements from the bow triggering melody notes. 

To generate harmony from the rhythms we used a uni-directional auto-encoder, modeled on Piano Genie \cite{donahue2019piano}. The generated rhythms were assigned a number between 0 and 7, which corresponded with low to high pitches. Melodies would always be in the range 4 to 7, with the accompanying parts 0-3. This allowed us to input any number of chord and melody notes, before the system return the pitches. The system is trained on data from International Piano-e-Competition, which includes a range of piano performances. Violin playback was done with and 8 Dio Orchestra Kontakt Library.

\section{System Integration}\label{sec:introduction}

\begin{figure}[h]
	\centering
		\includegraphics[width=\columnwidth]{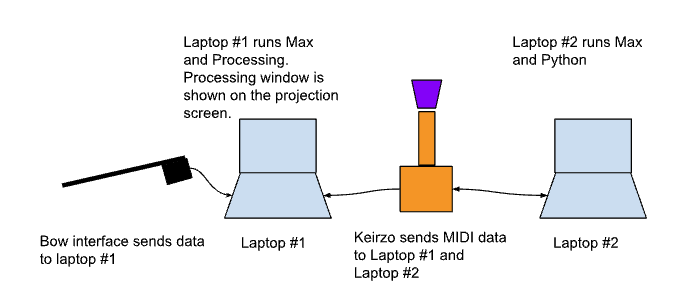}
	\caption{Communication Protocol}
	\label{fig:communication.png}
\end{figure}

Two computers were used to perform this work, one dedicated to the violin bow interface (Laptop 1, Figure~\ref{fig:communication.png} and the other to the synthesized string orchestra and Keirzo (Laptop 2, Figure~\ref{fig:communication.png}. Keirzo's drum hits were also sent to Laptop 1.
Data from the interface was parsed into Max/MSP using the serial port object and was received directly from the Zigbee receiver module. Max then communicated necessary data to the Processing sketch using the Open Sound Control protocol. Processing was able to receive information from Max via the external OscP5 library\footnote{\href{https://sojamo.de/libraries/oscp5/}{OscP5 for Processing}}.
Keirzo's drum hits were sent into the same Max patch as MIDI values using serial communication via Wifi, using a dedicated router.

Within this work, each sensor was specifically mapped to musical and visual parameters. The slider influenced the speed of Keirzo's drumming and synthesised strings. The buttons were mapped to change modes of performance and density of orchestral texture, controlling levels of intensity within each mode. These mappings gave the violinist a degree of interactive control and allowed them to structure the musical composition harmonically and according to timbral variance in real-time. The acceleromter provided values for both, the visuals and the synthesised orchestral strings, which were designed to be reactive but not controlled.

\section{Practice-Led Perspective}\label{sec:introduction}
Following the development we aimed to critically engage with the intricacies of designing, implementing, and interacting with new musical interfaces and robotic musicianship. This section aims to unpack issues and concepts that arose from our artistic intention, technical challenges, and aesthetic considerations that underpin the development and realization of the work.

\subsection{Acoustic vs. Synthesized Sound: Embodiment and Disembodiment}
A common consideration within robotic musicianship is the role of acoustic physical sound that is embodied and from the robot, compared to sounds emanating from a speaker and disembodied from any physical source \cite{savery2021shimon}. This embodiment adds a tangible, visceral layer to the performance, as audiences can visually connect the sound source with its auditory output. In contrast, the synthesized string orchestra, while rich in tonal quality, lacks a physical form or visual counterpart. 

This performance created a stark contrast between the embodied acoustic sounds of the drums and violin and the disembodied nature of the synthesized string orchestra. A continual concern of the creators was what agency would be perceived from the synthesized string orchestra, and whether it would be perceived as part of the robot or separately. For this reason the drumming was synchronized to the string sounds, through almost complete rhythmic unison.In the future, the contrast between embodied and disembodied sounds offers unique opportunities for creative expression, with opportunity to explore audience expectation, and variation in synchronization.

\subsection{Playing with Artificial Agent}
The process of developing this work came about in three stages - technical development (mapping interface sensor values, mapping Keirzo's MIDI values), rehearsing (learning how to use the mappings, getting used to playing with Keirzo) and finally, performing the piece. The initial phase opened up new creative possibilities to design approaches of interacting with an Artificial Agent as a playful performance partner. The second stage was the most challenging due to its high cognitive load on the violinist, as they learned to play the bow interface with its new mappings, develop the structure of the performance and choreograph  interactions with a robotic musician in a meaningful way.

Playing with an Artificial Agent challenged the violinist both physically and creatively. The compositional requirement of controlling a performance partner through a musical interface required patience and practice. On the other hand, lacking control due to external parameters associated with the nature of this work also took some getting used to. For instance, not knowing the tonality of the string samples at the start of each performance iteration, or the novel sonic nature of controlling intensity and texture density rather than volume. Improvising alongside an Artificial Agent also meant developing a new listening approach, forcing the violinist to allow Keirzo's time to respond and contribute musically, whilst simultaneously driving the direction of the piece through interface manipulation and intuitive decisions. Throughout this process, the violinist had developed a creative relationship with Keirzo, which evolved into a new performance style and improvisation strategies, unique to this experience. 

The opportunities that present themselves within an experimental work such as this, where the technologies are evolving alongside the composition are vast and exciting. A piece of music where learning to use the technology is part of the composition process pushes the human artist to explore new creative approaches, extending their existing skill-set and vastly expanding their creative scope.

This collaboration was a robust learning experience and future work with Keirzo will benefit from the groundwork laid down in this composition. Exploring more intricate control parameters within the mapping solutions could result in a more cohesive composition structure, where the synthesized strings are more sympathetic to the acoustic audio. Developing Keirzo's drumming vocabulary would also enhance the human/robotic musician interactions with a richer pallet of sonic and rhythmic variance.

\subsection{Stage Layout}
Based on the physicality of the performers and the visual component, our staging choices were somewhat limited. The projection screen inside our occupied space was positioned against the back wall, meaning the violinist and Keirzo needed to be either directly in front or beside it. The violinist chose to be positioned stage right, so that sound holes of the instrument would project towards the audience. Having Keirzo positioned stage left, allowed for the two collaborators to watch each other whilst also being able to see the screen. Although the space was large, the positioning of the video camera created a slightly minimised representation, especially of the projection screen, which had a much grander presence within a live setting.

The triangular set-up used for this performance had its benefits from a performer perspective, allowing good collaborative visibility. However, the violinist felt somewhat restrained and awkward in their positioning and gestural freedom. This was primarily due to the positioning of the projection screen, as well as the physical strain of manipulating the interface for musical and visual interactivity and compositional structure. 

In future works, deeper consideration could be placed into staging lighting and wardrobe. A new experimental work that combines multiple elements that may be unfamiliar or uncomfortable for a general audience, could create a sense of ease through deliberate performer positioning as well as colour codes and projection surfaces. The human performer juxtaposed beside a robotic musician, taller and larger than them could appear as menacing or disconnected. This issue could be muted by experimentation in performer positioning as well as colour schemes in performer wardrobe against the robotic musician colour schemes. The sterile placement of the projection screen could also be improved by incorporating softer fabrics such as silk scrims or bleeding the projection onto adjacent surfaces.
As the audience looks to decipher meaning within an unfamiliar setting, making conscious staging decisions could help communicate the intentions of the creative artists, building a more inclusive performance environment.

\subsection{Visual Design Strategies}
The visual element of this work was intended to be an augmentation of the physical performance acting as an extension of the performer's gestures whilst aiding in communicating the meaning behind the experiment. Each component of the visual design was tied in closely with the music, morphing with the violinist's gestures and changes in sonic modalities. As the violinist manipulated the interface, the data would simultaneously affect the synthesised sound and the projected visuals. A button pressed changed the colour of the particle system and differing slider values changed the wind force.

With a strong physical onstage presence from the violinist and Keirzo, designing visuals that would complement the performance was difficult, raising questions about how the visuals could be effectively incorporated into the piece as an integral element rather than a background distraction. One solution was to derive colours from Keirzo's design as well as visualising the drum hits as moving lines, which resemble drumsticks and spatially organised in correlation to their pitch, with hi-hat positioned top center, bass drum, bottom center and the ride and tom, vertically on either side. The violinist's bow gestures would then compliment Keirzo's drumming with a softer looking particle system that changed its appearance in response to bow strokes and bow speed. The three bouncing ellipses were the only independent element of the visual design, aimed to offset the direct drumming and bow gesture translation whilst creating a playfulness that the work inherently possesses. 

Future works for Keirzo and augmented violin will benifit from exploring visuals that contribute to the narrative development of the work, using more sophisticated mapping solutions, such as machine learning and genetic algorithms. The visuals in this project were overly simplistic and could have been absent. If visuals are to be incorporated within this setting, they would need to be an equal contributor to the piece, acting as a third collaborator.

\section{Audience-Led Perspective}\label{sec:introduction}
In the course of our research we conducted an exploratory online study aimed at gathering a broad spectrum of views on the performance. The primary objective was to ascertain the general audience's perceptions and to contrast these with our observations, particularly focusing on the variance in issues raised by both parties. We intentionally aimed to present the work without program notes or context, to attempt to contextualize broader opinions on the work without first directing attention or focus, or biasing the audience in anyway. 

To facilitate this study, we recruited 50 participants through Prolific, a platform known for academic and market research recruitment. These participants were exposed to a four-minute condensed version of the performance, which was made accessible online. Following the viewing, participants were prompted to respond to two open-ended questions. The first question sought their initial perspective on the performance, while the second question, visible only after the first had been answered, aimed to understand the perceived role of technology in the project.

Compensation for participation was set at £2 per participant, with an incentive structure in place to encourage detailed and insightful responses. The top 25 responses, selected based on their attention to detail and the novelty of their perspectives, whether positive or negative, were awarded an additional £1.50. Furthermore, the top 5 responses received an increased bonus of £3.00. Of the 50 individuals recruited for the study, 6 were disqualified due to incomplete video viewing, submission of brief or irrelevant responses, or failure to engage with the content meaningfully. Consequently, the study's final participant count stood at 44 individuals. The demographic breakdown of these participants included 30 women, 13 men, and 1 non-binary individual. The statistical analysis of participant demographics revealed a median age of 32 years, with a standard deviation of 11.92 years and an age range spanning from 21 to 75 years.

\subsection{Latent Dirichlet Allocation}

Thematic analysis of viewer feedback on a robot-violinist collaboration video was conducted using Latent Dirichlet Allocation (LDA), a topic modeling approach that identifies topics within documents. The analysis utilized the \texttt{sklearn} library's functions for pre-processing and topic modeling \cite{pedregosa2011scikit}. Data was processed to remove common stopwords, facilitated by \texttt{CountVectorizer}. This step produced a document-term matrix representing term frequencies across responses. LDA was then applied to this matrix, configured to extract five topics, revealing the distribution of words significant to each theme.

The LDA model identified five key themes, with the topic names added by the authors:
\begin{enumerate}
    \item \textbf{Musical Performance}: Discussions centered on the artistry and performance quality.
    \item \textbf{Human-Robot Interaction}: Insights into the collaborative dynamics between the human and robot.
    \item \textbf{Technological Enhancement}: The role of technology in augmenting the performance experience.
    \item \textbf{Emotional Response}: Viewer emotional reactions to the performance.
    \item \textbf{Technical Execution}: Comments on the technical aspects of the performance and robot's involvement.
\end{enumerate}

\subsection{Audience Themes}
In addition to the algorithmic theme analysis we hand-coded the interviews and found multiple reoccurring themes. At times these themes matched our own concerns while also raising new issues. 

\subsubsection{New Technology and Musical Aesthetics}
The integration of technology into musical performance often introduces audiences to an unprecedented artistic experiences. Viewers noted the uniqueness of witnessing technology play an instrument, marking a departure from traditional musical performances. Critiques of the musical output highlighted a divergence from conventional musical tastes, with some viewers finding the combination of drums and violin, mediated by technology, abstract and unfamiliar. One participant noted:``My initial thoughts are that I don't understand or appreciate the music that was performed. I like both classical music and pop music in general but that was just a bit too abstract." Many participants however said they enjoyed hearing a new approach to music and that it was an interesting role for the technology to help create new approaches to music. 

We also believe the work may have benefited from a more gradual introduction to each piece of technology in the work. Multiple participants noted an initial confusion, such as ``I watched the video trying to make sense of all the moving parts -- the violinist, the drumming robot and the screen. I could not figure out how these were interrelating and influencing each other. It seems that the violinist was improvising off of the drums and the orchestral music, but not the screen. I couldn't figure out if the robot was making all of the non-violin music, or just the beats. In general, I liked the improvised music -- reminded me of jazz.' In future work gradual introduction and education about the roles of each technology could improve how audiences unfamiliar with the work understand the technology.

\subsubsection{Anthropomorphize}
The design of the robot elicited primarily positive reactions, however there were mixed reactions to the human-like elements. While some found the humanoid gestures of the robot endearing, others perceived it as deceptive, preferring a clear distinction between the robotic and human elements. One participant described ``Had they not tried to anthropomorphize the robot, I would have enjoyed it more. It would have seemed somehow less deceptive to have it presented as a machine without any attempt to personify it.'' This dichotomy raises questions about the role of anthropomorphism in technology, particularly in artistic contexts, and its impact on audience reception.

\subsubsection{Interaction}
Viewer responses varied on the perceived interaction between the musician and the robot, with many praising the interaction, such as ``I thought the video was very interesting because it was clear that the musician and robot were interacting'' and ``it felt like a performance given by a duo''. Others felt the robot was too dominating and ``not complementing the performer well'', perhaps suggesting they believed the robot should play more of an accompanying role. 

\subsubsection{Robotic Soul}
The essence of human creativity emerged as a central theme, with some viewers commenting on the robot's inability to replicate the soulfulness and emotional depth characteristic of human performances. Quotes included ``It lacked the soul and flavor a human would produce when playing an instrument.'' and ``It felt like it was lacking something, and that was probably some emotion.''. Other participants noted that ``A robot making music isn’t as appealing as a person.'' and ``In my view, humans are best at creating art''. Nevertheless the vast majority of participants treated the robot as another collaborator on stage.

\subsubsection{Ethical Concerns}
Although not predominant, some comments touched on ethical considerations related to the use of musicians' data in training the robot, highlighting concerns over consent and compensation. One participant wrote ``My first thought was to wonder if the actual musicians whose data had been used to train the robot had consented. I then wondered if they were at least compensated''. This was not a key concern however, and almost all participants instead focused on the end product.

\subsubsection{Robot Agency}
Questions about the robot's autonomy and the extent of its pre-programmed versus live-responsive capabilities indicated a curiosity about the technology's sophistication. This interest in robot agency leads to a broader discussion on the evolving role of artificial intelligence in live performances, examining the balance between programmed routines and adaptive, interactive behaviors. Participants asked ``was someone controlling the robot or did the robot control itself.'', ``It is hard to tell how much of the music playing was pre-programmed for the Robot or if it was able to sync up with the lady live on the stage.'' and ``Was it randomly responding or doing something pre-programmed''.



\begin{figure}[htbp]
\centering
\includegraphics[width=\linewidth]{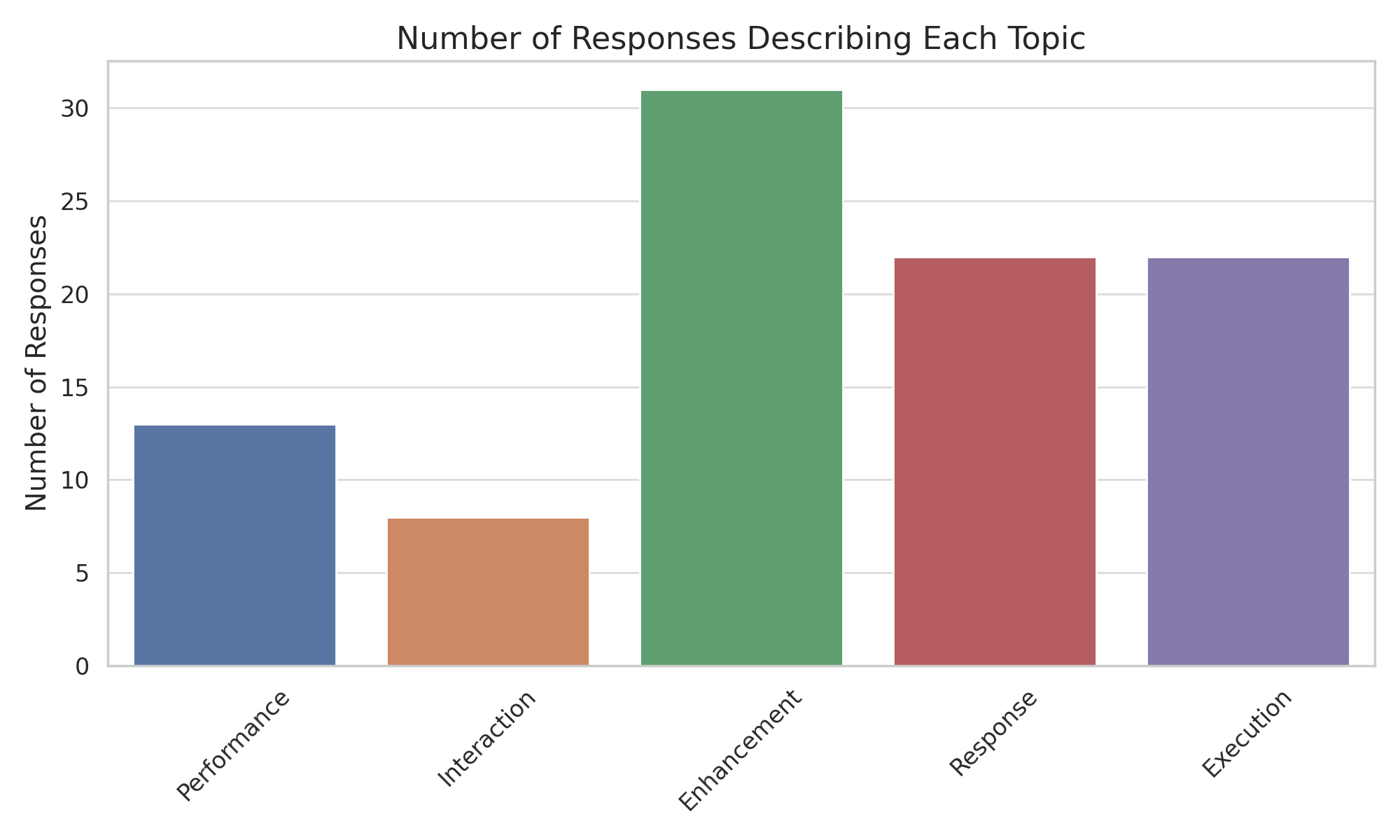}
\caption{Distribution of viewer feedback across identified themes.}
\label{fig:viewer_feedback_themes}
\end{figure}

\section{Discussion, Future Work and Conclusion}\label{sec:introduction}
In this paper, we discussed the process of creating a mixed-media, experimental work for augmented violin, robotic musician and interactive visuals. We also presented discussion points around themes that we, as developers of this project were most concerned with. Our perspectives are grounded in our own personal relationships with the technologies used within this project as well as external constraints that shape the final outcome of the work. These include time, existing technology, available resources such as performance spaces and documentation tools and our own limitations as practice-based researchers and creative artists.

Although driven to create a musically and visually engaging work, our primary concerns are associated with the development processes of this and future works, shaped by ongoing personal and external evaluation. Having lived with this piece from its conception, it can be difficult to ascertain a removed, holistic view of the final product.

The audience perspectives as presented in the survey were derived from a point of unfamiliarity with the project and the music, arming them with a fresh perspective. Although using slightly different terminology, it is important to note, that much of the discussion points brought up by the participants aligned with our own concerns.  For instance, \textit{Musical Performance}, although broad, relates back to both, embodiment/disembodiment of sound and the human-computer interaction approaches as seen by the violinist. Concerns raised about anthropomorphism could also be linked to the strategies taken by the violinist in interacting with Keirzo and shaping the structure of the piece.

In future works, placing greater significance on audience accessibility through points of departure surrounding familiarity of style and introduction to the technologies being used will decrease audience confusion and feelings of alienation.  Transparency in data being used for training models will also aid in softening feelings of hostility. Overall, we feel optimistic that as our technologies develop alongside our composition strategies, there is robust potential for engaging and novel works incorporating robotic, virtual and human collaborators.

\bibliography{icmc2024template}

\end{document}